\documentclass{webofc}
\usepackage[varg]{txfonts}   
\begin{document}
\vspace*{-2cm}
\title{What ultracold atoms tell us about the real-time dynamics
of QCD in extreme conditions
\vspace*{-0.2cm}}

\author{\firstname{Jürgen} \lastname{Berges}\inst{1}\fnsep\thanks{\email{berges@thphys.uni-heidelberg.de}, Quark Matter 2023 plenary https://indi.to/VPVMX
} 
}

\institute{Heidelberg University, Institute for Theoretical Physics, Philosophenweg 16, 69120 Heidelberg
}

\abstract{%
I review developments of how compact table-top setups with ultracold atoms can help us to understand the more complex real-time dynamics of QCD probed in heavy-ion collision experiments. 
\vspace*{-0.6cm}}
\maketitle
\section{Simulating quantum systems by other quantum systems}
\label{intro}
The spatio-temporal evolution of heavy-ion collisions is often put into context with the evolution of the early universe. Despite differences, there are questions about their dynamics that can be studied in both systems alike. In recent years another type of system comes into play: setups with ultracold quantum gases provide versatile model systems for simulating related dynamical questions that cannot be simulated on present-day classical computers. 

What makes cold-atom setups quantum simulators is their very high experimental control~\cite{Bloch08}. To some extent one can design Hamiltonians with any desired properties by also applying electromagnetic fields. With these one can modify the dimensionality of space (e.g.\ by optical traps or by coupling internal "synthetic" dimensions), symmetries, field content and interaction strength. Together with well-controlled initial state preparation, simultaneous momentum and space imaging, and the ability to monitor the detailed history of the time evolution these setups can be very powerful tools. Of course, they can never replace heavy-ion experiments probing fundamental properties of QCD in nature, but quantum simulations can be extremely helpful in understanding them.  

\begin{figure}[t]
\centering
\includegraphics[width=13cm,clip]{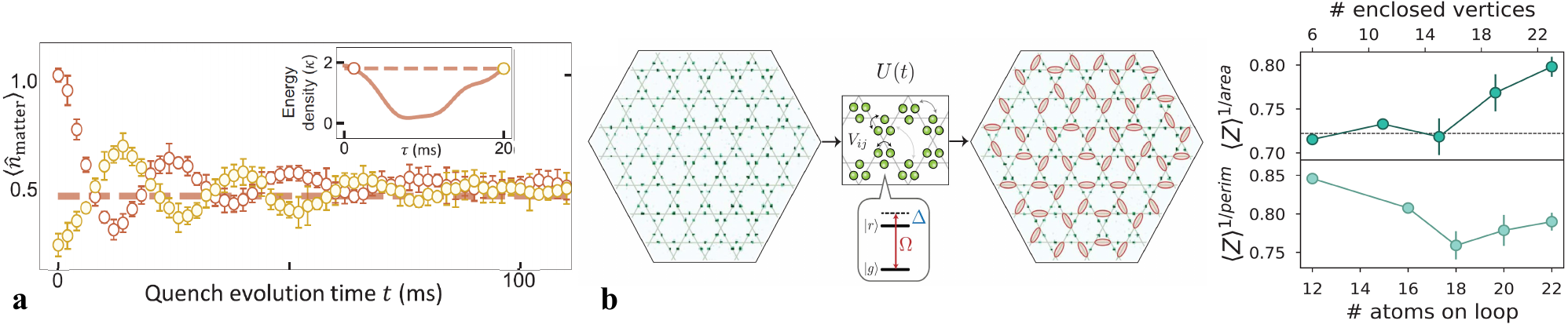}
\vspace*{-0.6cm}
\caption{\textbf{a)} Quantum simulation of the thermalization dynamics for an Abelian gauge theory in 1+1 space-time dimensions using ultracold atoms. Evolutions from different far-from-equilibrium initial conditions with same energy density approach the thermal value at late times. Fig.~adapted from~\cite{Zhou22}. \textbf{b)}~Effective realization of a discrete ($Z_2$) gauge symmetry in 2+1 space-time dimensions using Rydberg atoms placed on the links of a kagome lattice. Measurements of non-local observables indicate an approximate area/perimeter law scaling depending on the loop size. Figs.~adapted from~\cite{Semeghini21}.}
\label{fig-gauge}       
\vspace*{-0.6cm}
\end{figure}

The ultimate goal is the engineering of model Hamiltonians for gauge fields as closely as possible to those relevant for heavy-ion collisions, but there is still some way to go. For instance, Fig.~\ref{fig-gauge}a shows results from a quantum simulation of the thermalization dynamics for a simple Abelian gauge theory starting from different far-from-equilibrium initial conditions as a function of time, which is notoriously difficult on classical computers~\cite{Zhou22}. While many state-of-the-art examples concern 1+1 space-time dimensions, there is also exciting progress towards higher dimensional implementations. Fig.~\ref{fig-gauge}b illustrates the effective realization of a discrete gauge symmetry in 2+1 space-time dimensions where non-local observables are measured, identifying an area-law behavior reminiscent of what one would like to do to detect confinement in QCD~\cite{Semeghini21}. For a recent review see Ref.~\cite{Bauer23}.   

In view of the complexity of heavy-ion collisions these are still relatively simple examples, and to see how one can make even further progress already with present-day ressources it is helpful to distinguish analog and digital quantum simulations. Digital quantum simulations decompose the desired Hamiltonian into smaller pieces or "gates" by breaking up or "Trotterizing" the unitary time evolution. The aim is to encode and read out as much as possible of the microscopic quantum state. This highly flexible and demanding approach underlies also programmable quantum computers. 

Instead, in this short review I concentrate on analog quantum simulation for which large-scale systems with excellent coherence properties are already available in the laboratory. Analog simulations exploit similarities between quantum systems such as symmetries. Even universal behavior can occur, where microscopic details don‘t matter. This is well known for continuous phase transitions in equilibrium, such as for a QCD critical point that shares exact quantitative properties with critical opalescence in water. New developments reveal that there can be a remarkable degree of universality even far from equilibrium, also away from any phase transition, which is a tremendous simplification for quantum simulations! 

\vspace*{-0.1cm}

\section{Pre-equilibrium dynamics: Universality far from equilibrium}
\label{universal}

\begin{figure}[t]
\centering
\includegraphics[width=13cm,clip]{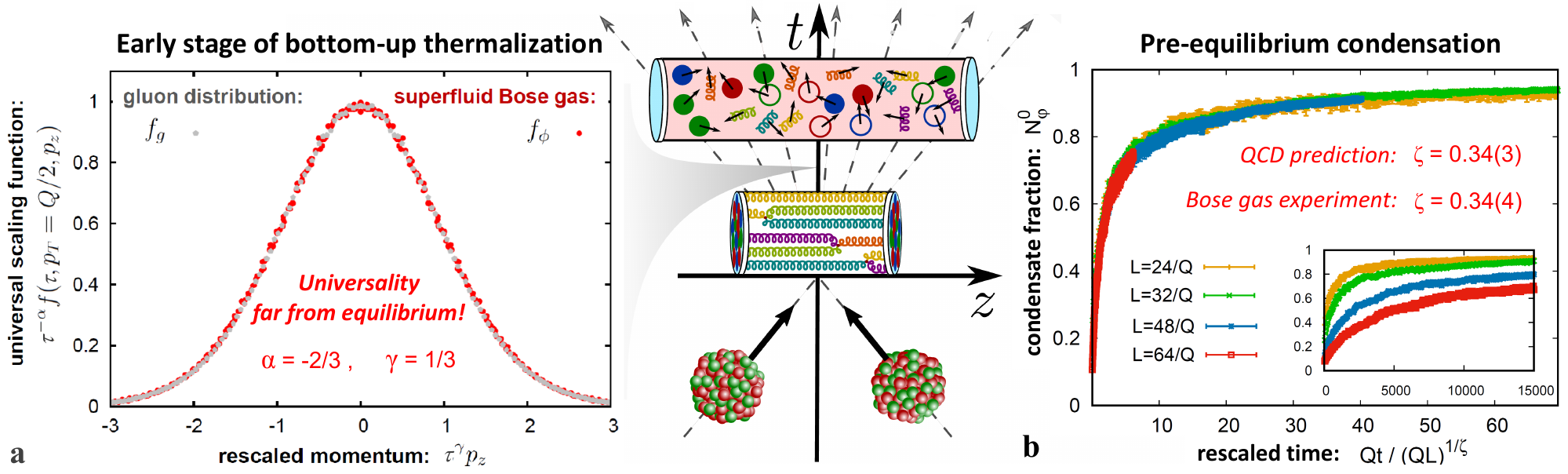}
\vspace*{-0.6cm}
\caption{\textbf{a)} Schematic evolution of the quark-gluon plasma in a heavy-ion collision. Attractors during the pre-equilibrium stages lead to universal behaviour characterized by scaling exponents $\alpha$ and $\gamma$ of the bottom-up thermalization scenario. The left figure shows the precise agreement of the predicted early-time dynamics of gluons and scalars (Bose gas) at different proper times $\tau$ and momenta $p_z$ for longitudinally expanding systems, all collapsing to a universal rescaled curve for same $\alpha =-2/3$ and $\gamma = 1/3$. Figs.~adapted from~\cite{Berges15,Berges21}. \textbf{b)} Analog quantum simulations with ultracold atoms also exhibit pre-equilibrium condensation in this regime. After identifying the gauge-invariant observable for condensation in QCD, a corresponding phenomenon is predicted for gluons as displayed in the figure~\cite{Berges23}. The exponent $\zeta$ for condensation captures the finite-size scaling with system size $L$ or the infrared behavior for the non-expanding systems, showing quantitative agreement between Bose gas experiments and QCD prediction. The experimental value $\zeta = 0.34(4)$ is taken from~\cite{Glidden21}.
}
\label{fig-universal}       
\vspace*{-0.6cm}
\end{figure}

Universality out of equilibrium implies insensitivity of the dynamics to details of initial conditions. Predictions of attractors in the evolution of the quark-gluon plasma exhibit such an effective loss of information already far from equilibrium at rather early stages. The middle graph of Fig.~\ref{fig-universal} illustrates the schematic evolution where two colliding nuclei at sufficiently high energies leave behind predominantly gluons~\cite{AM}. Remarkably, if one varies the initial gluon occupancy or the anisotropy of their distribution, then the subsequent evolution quickly becomes insensitive to where it started. This nonthermal attractor behavior in the highly occupied regime at early times~\cite{BBSV14} singles out the bottom-up thermalization scenario~\cite{BMSS01}. A somewhat similar situation is encountered also for lower occupancies and stronger couplings, which emerge as the system further expands. For strong couplings this has been pointed out for holographic conformal field theory using many different initial states, all converging to a hydrodynamic attractor~\cite{Heller12}. For recent reviews see Refs.~\cite{Jankowski23,Berges21,Schlichting19}.  

For quantum simulations of the far-from-equilibrium dynamics with ultracold atoms, one first has to set up a dictionary~\cite{Berges15}. I will concentrate on sufficiently energetic heavy-ion collisions such that the scale-dependent gauge coupling $\alpha_s(Q_s)$ is small at the characteristic saturation scale $Q_s$. On the cold-atom side this corresponds to a small diluteness parameter $\chi = \sqrt{n a^3}$ being a dimensionless combination of the scattering length $a$ and atom density $n$. These two parameters also determine the inverse coherence length $Q = \sqrt{16 \pi a n}$, which plays the role of the saturation scale. Despite the weak coupling parameters, because of high characteristic occupancies $\sim 1/\alpha_s$ and $\sim 1/\chi$ respectively, the dynamics evolves from a strongly correlated initial state that is over-occupied compared to equilibrium~\cite{Cosmo}.

Comparing the predicted pre-equilibrium dynamics of these systems from classical-statistical simulations, one observes a striking degree of universality far from equilibrium: Instead of different gluon and (relativistic) Bose gas distribution functions, a single universal scaling form emerges as shown on the left of Fig.~\ref{fig-universal}a~\cite{Berges15,Berges21}. This represents a strong restriction for the dynamics, since the nonequilibrium evolution in this regime is encoded in a time-independent scaling function that only depends on the product of time and momenta with universal scaling exponents for the longitudinally expanding systems with transverse and longitudinal momenta. The figure shows the precise agreement for the gluons and the Bose gas at different times and momenta, which is a remarkable manifestation of the effective loss of information about microscopic details already at an early far-from-equilibrium stage.

While the full quantum dynamics of these nonequilibrium systems cannot be simulated on classical computers, we can inquire cold-atom simulators in the laboratory. Universal self-similar scaling due to the nonthermal attractor has been experimentally observed first in elongated traps~\cite{Prufer18}. More recent Bose gas results in 3+1 space-time dimensions give a very detailed picture of the underlying dynamics without expansion~\cite{Glidden21}. One experimentally observes a self-similar energy flow towards higher momenta, which can be rescaled to collapse to a single universal shape as expected. At the same time, the cold atom system exhibits  the striking phenomenon of pre-equilibrium condensation! A self-similar particle flow towards low momenta arises, i.e.\ in the opposite direction than the energy flow towards higher momenta. Due to universality this dynamical build-up of a macroscopic occupation number in the infrared is insensitive to system parameters such as the interaction strength~\cite{CondensationDualCascade}.

It took some time to sort out what is a suitable gauge-invariant observable for condensation in the non-perturbative low-momentum regime of QCD~\cite{Berges23,Berges20}. Fig.~\ref{fig-universal}b shows the predicted build-up of the condensate based on the gauge-invariant phase eigenvalue of a spatial Polyakov loop, which is a variant of the observable used to signal the deconfinement transition in equilibrium. The extracted universal exponent $\zeta$ for condensation captures the finite-size scaling with system size, or the corresponding infrared behavior, and the inset of Fig.~\ref{fig-universal}b shows the results as a function of time without finite-size rescaling. The quantitative agreement between the results for $\zeta$ from the Bose gas experiment and the QCD prediction is intriguing and demands further understanding of the relevant infrared effective theory.

There is much progress in our understanding of how the rapid effective loss of information about microscopic parameters and initial conditions occurs, and the related question about the emergence of effective theories such as hydrodynamics far from equilibrium both in QCD and cold atoms. The advances include identifying time-dependent pre-scaling exponents as slow effective hydrodynamic variables~\cite{Mazeliauskas19}, or the adiabatic hydrodynamization framework~\cite{Brewer22} for the over-occupied plasma at early times with close similarities to the discussions of hydrodynamic attractors for stronger couplings at lower occupancies~\cite{Alalawi22}.

\vspace*{-0.1cm}

\section{Origin of collective behavior in small systems}
\label{collective}

\vspace*{-0.3cm}

\begin{figure}[t]
\centering
\includegraphics[width=13cm,clip]{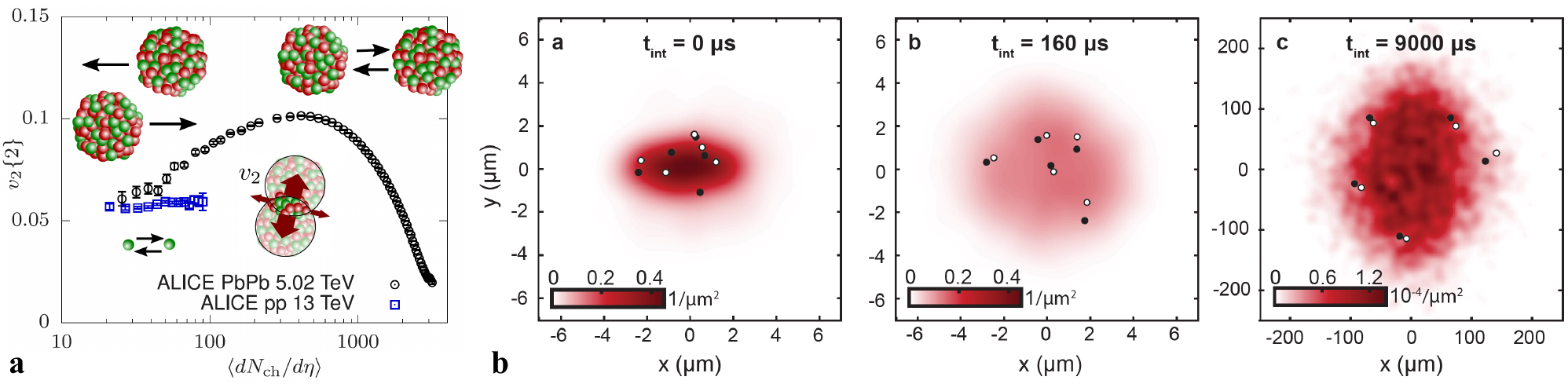}
\vspace*{-0.6cm}
\caption{\textbf{a)} Illustration of collective behavior indicated by elliptic flow measurements from heavy-ion and proton-proton collisions versus the charged-particle multiplicity density. Fig.~adapted from~\cite{AM}.
\textbf{b)}~Emergence of pairing and collectivity in few Fermion quantum gases. Pairing of 5+5 strongly interacting spin up and down atoms is illustrated by emerging pairs of black and white dots during anisotropic expansion for a single realisation. Sampling over experimental realisations is used to obtain their distribution. The aspect ratio of the horizontal and vertical extensions of the distribution becomes inverted due to interactions indicating hydrodynamic flow. Fig.\ adapted from~\cite{Brandstetter23}. 
}
\label{fig-collective}       
\vspace*{-0.6cm}
\end{figure}

Nonequilibrium aspects play also an increasing role when going to smaller systems such as arising from proton-nucleus or even proton-proton collisions, where observations of collectivity challenge our current understanding of the underlying dynamics~\cite{Collectivity}. In the following, I will highlight a set of very inspiring cold-atom experiments concerning the emergence of superfluid hydrodynamics and the origin of collective behavior in small systems.   

An important finding from the RHIC and LHC experiments is that the quark-gluon plasma behaves as a nearly perfect fluid with a small shear viscosity to entropy density ratio. The only competitor is a strongly coupled ultracold Fermi gas, and observations such as hydrodynamic flow have been demonstrated for both systems alike~\cite{Adams12}. Fig.~\ref{fig-collective}a illustrates elliptic flow measurements from heavy-ion as well as proton-proton collisions at the LHC, which reflect significant angular modulations in the number of produced particles. The phenomenon of collective particle emission is typically associated with the formation of a hydrodynamic medium based on a separation of scales between micro- and macro-physics. However, for small systems with only few tens of final-state particles the relevant length scales such as system size, inter-particle spacing, and mean free path can become comparable.

Here ultracold quantum gas setups with resolution capabilities even down to single particle level, and deterministic control over particle number and interaction strength, allow one to explore the boundaries between a microscopic
description and a hydrodynamic framework in unprecedented detail. In the past, in Ref.~\cite{Thomas02} the anisotropic expansion of a macroscopic number ($\sim \!\! 10^5$) of Lithium atoms was investigated, and the aspect ratio of the horizontal and vertical extensions of the cloud was seen to become inverted due to interactions indicating hydrodynamic flow. It has been pointed out for these experiments that a superfluid component seems important for their hydrodynamic description.

Strikingly, recent experiments with few atoms show similar results with the emergence of hydrodynamic flow already for small systems~\cite{Brandstetter23}. Fig.~\ref{fig-collective}b shows the positions of 5+5 strongly interacting spin up and down atoms at different times, and sampled over many experimental realizations of the same quantum state to obtain their distribution. For the interacting system, one always observes an inversion of the aspect ratio at all atom numbers. Moreover, pairing of spin up and down atoms is seen to arise explicitly, and it is an exciting question to further investigate to what degree a superfluid-like behavior plays a role for the origin of collectivity in these small systems.  

\section{Outlook}
\label{outlook}

So far, the state-of-the-art bottom-up thermalization scenario for the quark-gluon plasma does not include collective dynamical phenomena, such as pre-equilibrium condensation. On the phenomenological side, there seem to be increased deviations between data and state-of-the-art model predictions towards low-momentum (pion) yields
\cite{Devetak20}, and it would be striking to establish if this can be traced to infrared collective effects. Close to equilibrium there exists a well-established framework for the description of collective phenomena with condensates or order-parameter fields~\cite{Halperin}, and a related example is the enhancement of produced low-momentum pions from QCD chiral critical fluctuations~\cite{Grossi}. However, our understanding of effective theories far from equilibrium with order-parameter fields for condensation or superfluidity is still in its infancies, and quantum simulations have a tremendous potential to help us understanding the complex real-time dynamics. 

Of course, there are many more important topics in this context which I didn‘t mention. Further themes include hard probes or, more generally, open system dynamics in heavy-ion collisions~\cite{OpenSystems}. Some of the universal aspects of highly energetic jets loosing energy to the quark-gluon plasma is, e.g., captured in terms of a mini-jet attractor appearing in the last stage of the bottom-up thermalization scenario~\cite{Schlichting19}. The analogue situation to hard probes in cold atoms is reminiscent of impurity dynamics/Polarons or fastly moving excitations in driven Bose-Einstein condensates~\cite{Cetina}. 

A prime discipline of cold-atom experiments is also entanglement detection and entropy for the discussion of thermalization~\cite{Entanglement}. For instance, taking into account that only the causal regions determined by the particle and detector horizons are observable, then tracing out the acausal parts leads to an effective temperature for the reduced density matrix of the vacuum system~\cite{BFV}. For the Schwinger model for hadronization a time-dependent temperature reminiscent of the Hawking temperature but without any acceleration emerges from entanglement in this way, which can play an important role for small systems that break up early. Advances in space-time resolved entanglement detection capabilities for ultracold quantum gases make this a very promising future research direction for quantum simulations~\cite{Kunkel22}.\\

\textit{Supported by
the DFG under the Collaborative Research Center SFB
1225 ISOQUANT (Project-ID 27381115) and the Heidelberg
STRUCTURES Excellence Cluster under Germany’s
Excellence Strategy EXC 2181/1-390900948.}

\end{document}